# Investigation of Vibrational Frequency of Canine Vocal Folds Using a Two-Way Fluid-Solid Interaction Analysis

*Abolfazl Mohammadi Gorjaei[1], Mohammad Ali Nazari[1], Asghar Afshari[1], Saeed Farzad-Mohajeri[2], Pascal Perrier[3]*

[1] School of Mechanical Engineering, University of Tehran, Tehran, Iran
[2] Department of Surgery and Radiology, Faculty of Veterinary Medicine, University of Tehran, Tehran, Iran
[3] Univ. Grenoble Alpes, CNRS, Grenoble INP, GIPSA-lab, 38000 Grenoble, France

`amohammadig77@ut.ac.ir`, `manazari@ut.ac.ir`, `afsharia@ut.ac.ir`, `saeedfarzad@ut.ac.ir`, `Pascal.Perrier@grenoble-inp.fr`

**Introduction.** Speech is an integral component of human communication, requiring the coordinated efforts of various organs to produce sound (Titze & Alipour, 2006). The glottis region, a key player in voice production, assumes a crucial role in this intricate process. As air, emanating from the lungs in a confined space, interacts with the vocal folds (VFs) within the human body, it gives rise to the creation of voice (Alipour & Vigmostad, 2012). Understanding the mechanical intricacies of this process is very important. Studying VFs in vivo situations is hard work. However, the orientation, shape and size of VFs fibers have been extracted with synchrotron X-ray microtomography. (Bailly *et al*., 2018)

The investigation of mechanical properties of both human and animal VFs has been carried out through various methodologies in the literature. The mechanical properties of VFs have been studied using the uniaxial extension test (Alipour & Vigmostad, 2012) assuming a linear behavior, while the nonlinearity and anisotropy of VFs has been determined using a multiscale method as in Miri *et al*. (2013). Pipette aspiration has also been used to extract in vivo elastic properties of VFs (Scheible *et al*., 2023). Mechanical behavior of VFs layers in tension, compression and shear has been studied. (Cochereau *et al*., 2020). Fluid-structure interaction (FSI) simulations provide a valuable tool to gain a deeper understanding of voice production (Ghorbani *et al*. 2022). These simulations allow us to model the dynamic interplay between the VFs and air. Our research focuses on investigating the mechanical properties of canine vocal folds and utilizing these findings in an FSI simulation. Through this simulation, we aim to unravel how these mechanical properties affect voice production.

**Methods.** To investigate the mechanical properties of canine VFs, an in vitro study was conducted involving 6 mixed-breed dogs. The samples were harvested from canine cadavers euthanized for reasons unrelated to this study. In the following, the VFs were harvested and tested upon 3-4 hours post-animal sacrifice.

Experimental trials were carried out using the STM-1 device (SANTAM Co.), equipped with a 100 kg load cell. Seven uniaxial tensile tests were done on each sample, with displacement rates of 1, 5, 10, 20, 40, 60, and 120 mm/min. The very slow rate of 1 mm/min was chosen to assess only elastic properties eliminating viscosity effects. Various hyperelastic models were used to fit the experimental data. Subsequently, for each model, both the mean and standard deviation (SD) were determined for the hyperelastic model parameters and their residuals.

For FSI analysis we used a simplified laryngeal model as a hollow cylinder with a diameter of 50 mm and a thickness of 3 mm. The overall length of the larynx was set at 100 mm. The VFs were modeled as a circular disc with a small elliptical fissure in the midst of the cylinder section. Boundary conditions were established based on pressure differentials, with the inlet gauge pressure set at 1200 Pa and the relative pressure at the outlet set to 0. To account for the turbulent nature of airflow within the larynx, we employed the K-epsilon method to solve the motion differential equations in a two-way fluid-structure interaction simulation using ANSYS FLUENT 2021.

This approach enabled us to investigate how the acquired mechanical properties of canine vocal folds affect the FSI simulations during phonation, resulting in a more comprehensive understanding of their impact. To determine the vibrational frequency of VFs, we calculated the time it took to reach maximum displacement and then quadrupled this value to obtain the period of vibration.

**Results.** The Yeoh 2nd order model emerged as the most fitting choice with its strain energy density function:
$$\psi = C_{10}(I_1 - 3) + C_{20}(I_1 - 3)^2$$
where $C_{10}$ and $C_{20}$ represent the model parameters and $I_1$ denotes the first invariant of the Cauchy-Green strain tensor. The associated material constants derived from this model and averaged between all six canine samples were found as: $C_{10}$=195.8±139.7 kPa, $C_{20}$=6765.2±1469.4 kPa.

Stevens (2000) has expressed the deformation of VFs in 8 steps. The vibration of the simplified VFs in our simulations shows the first four steps and thus indicates the half cycle of the VFs vibration. Therefore, the time duration of oscillating behavior of VFs in Y direction was used to compute the vibrational frequency of the simplified VFs. The vibrational frequency extracted from these results show a frequency about 125 Hz. This frequency has been calculated by inversing the twice the time taken for maximum displacement. Our findings indicate that the simplified larynx model, incorporating the extracted mechanical properties, displays a compelling behavior of voice production.

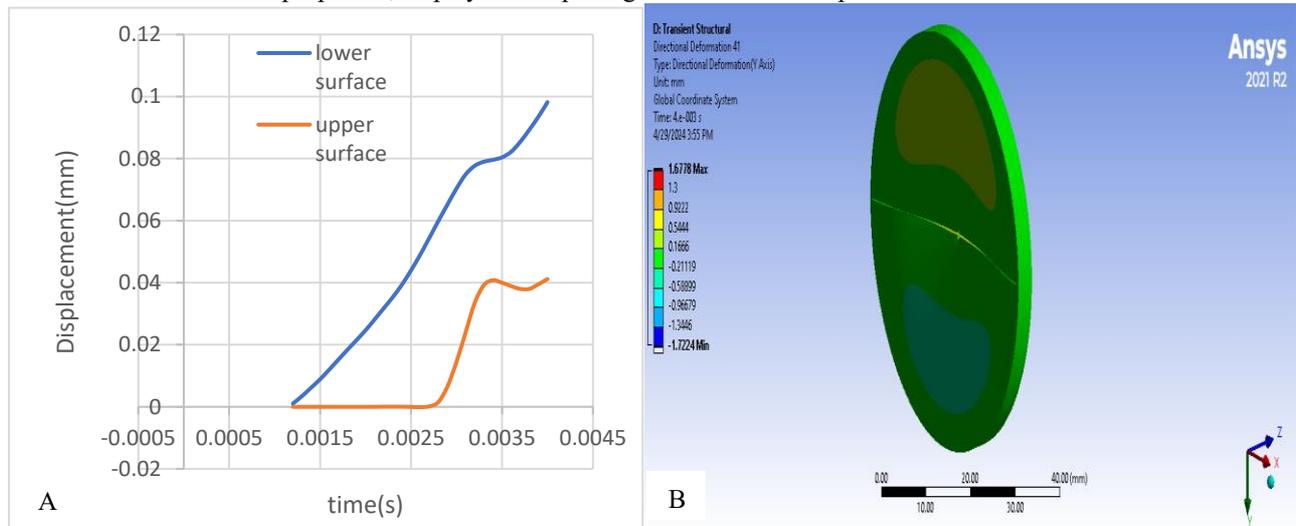

**Figure 1**: *The results of the FSI simulation A- Relation of the directional displacements of the VFs in Y axis over time for two points B- the deformed shape of VFs in time=0.004s*

**Figure 1-A**, (the left panel) shows the displacements in Y axis direction over time for two points in z direction on the surface of fissure located on upstream and downstream parts of model. As it can be seen the downstream point (blue curve) is opened while the upstream point (red curve) is still closed. In the following the upstream point is opened and the air flows through the opened fissure. **Figure 1-B** indicates the maximum deformation of the model when the contact between fissures opens. The frictionless contact has been defined in the surface of fissure (XZ plane).

**Discussion.** The mechanical properties of the VFs in the previous study of our group on speech production via FSI simulation were based on the cadaver studies by Alipour, F., & Vigmostad, S. (2012). A hyperelastic model (first order Ogden) was used to fit their experimental results (Ghorbani *et al.,* 2022). In the current research, the fresh canine VFs were used to have the VFs mechanical properties. The number of specimens and the control on strain rate in our tensile experiments gives promising behavior than the former one. At the same time, we are aware that the viscoelastic properties of VFs play an essential role in their behavior. The extracted experimental data of samples let us studying this behavior which is the subject of ongoing research.

Our results for the frequency of canine VFs are close to the results of Solomon et al. (1995) which reported the frequency of growl to be equal to 112 Hz. These results indicate that our simplified simulation and the derived mechanical properties exhibit promising accuracy in predicting vocal fold behavior.